\documentclass[runningheads]{llncs}
\usepackage{hyperref}
\usepackage{graphicx}

\begin{document}

\title{GeoLogic -- Graphical interactive theorem prover for Euclidean geometry}
%
%
\author{Miroslav Ol\v s\'ak\inst{1}\orcidID{0000-0002-9361-1921}}
\authorrunning{M. Ol\v s\'ak}
%
\institute{%
University of Innsbruck, Austria,\\
\email{mirek@olsak.net}}
\maketitle              
\begin{abstract}
Domain of mathematical logic in computers is dominated by
automated theorem provers (ATP) and interactive theorem provers (ITP).
Both of these are hard to  access by AI from the human-imitation
approach: ATPs often use human-unfriendly logical foundations while ITPs
are meant for formalizing existing proofs rather than problem solving.
We aim to create a simple human-friendly logical system for mathematical
problem solving. We picked the case study of Euclidean geometry as it
can be easily visualized, has simple logic, and yet potentially offers
many high-school problems of various difficulty levels. To make the
environment user friendly, we abandoned strict logic required by ITPs,
allowing to infer topological facts from pictures. We present our system
for Euclidean geometry, together with a graphical application GeoLogic,
similar to GeoGebra, which allows users to interactively study and prove
properties about the geometrical setup.
\keywords{Euclidean geometry  \and Logical system.}
\end{abstract}

\begin{figure}
  \centering
  \includegraphics[width = 9cm]{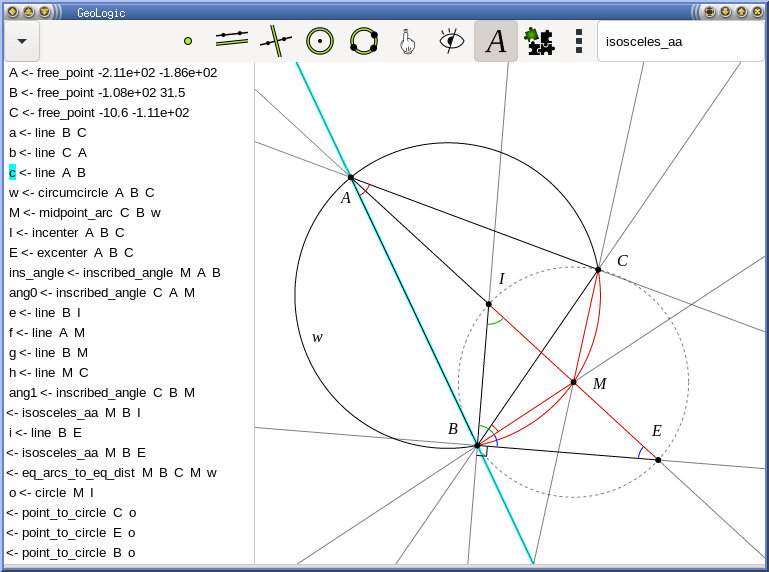}%
  \caption{GeoLogic screenshot}
\end{figure}

\section{Overview}

The article discusses GeoLogic 0.2 which can be downloaded from
\url{https://github.com/mirefek/geo_logic}. It is a logic system for
Euclidean geometry together with a graphical application capable of
automatic visualisation of basic facts (equal angles, equal distances,
point being on a line, ...) and allowing user interaction with the
logic system. GeoLogic can be used for proving many classical high
school geometry problems such as Simson's line, Pascal's theorem, or
some problems from International Mathematical Olympiad. Examples of
such proofs are available in the package. In this paper, we first
explain our motivation, then we give a description of the underlying
logical system, and finally we present and example of proving the
Simson's line to demonstrate GeoLogic's proving and visualisation
capabilities.

There are many mathematical competitions testing mathematical problem
solving capabilities of human beings, presumably most famous of which
is the International Mathematical Olympiad (IMO). Writing an automated
theorem prover (ATP) that could solve a large portion of IMO problems is a
challenge recognized in the field of artificial
intelligence~\cite{IMO-Challenge}, and could potentially lead to
strong ATPs in general.

IMO, as well as many regional mathematical olympiads divide problems
into four categories: algebra, geometry, combinatorics and number theory.
From a human solver's perspective, computer can significantly help
with solving geometry problems using an application such as GeoGebra --
it allows the user to draw the configuration precisely, and observe
how it changes when moving the initial points.

This is one of the reasons why we focused on geometry.
Our objective is to capture the steps performed by such human solver
in more detail, hoping it could eventually lead to better
understanding of human thinking in general.
Therefore, we are building an interactive
theorem prover, while preserving the usability as an exploration
tool. In future, we would like to experiment with machine learning
agents leading to human-like ATPs for geometry. Moreover, geometry is
concrete and visualisable. This allows to add computer vision
components to the future machine learning experiments.

Other advantage of reasoning in Euclidean geometry is that we don't
need complex logic. Most of geometrical reasoning involves only direct
proofs without higher order logic or case analysis. While some
geometrical proofs use case analysis for different topological
configurations, we use a different approach. We allow to infer
topological facts (such as orientation of a triangle) from the picture
(numerical model). This apparently proves only one case of the problem
(and its neighborhood), and could potentionally lead to
inconsistencies caused by numerical errors. However, we believe
inconsistency caused by a numerical error is unlikely because we
require the fact to be satisfied by a sufficient margin for postulating it,
Softening the logic so that it accepts a proof of just one
configuration is actually an advantage: It is a common case in
Euclidean geometry that a proof of a single configuration can be used
for proving the general case either by case analysis and analogies, or
by proving that the configuration is the only possible. Therefore,
introducing a flexible logic can be seen as providing an intermediate
step for finding a formal proof -- one first want to proof it in the
flexible logic of GeoLogic, and then to transform the GeoLogic's proof
into a formal one. Both of the subtasks are typically easier then the
original problem, and such proving procedure would reflect how a human
solver usually approach the problem.

Finally, even though our main motivation was not to make a pedagogical
tool, and we don't market GeoLogic as an application for an arbitrary
high school student in its current form, we also believe that GeoLogic
can be already interesting for talented students.
Our objective of making an user-friendly interactive theorem
prover for geometry is well-aligned with educational purposes, and if
it will get adopted in the future, it can help us with obtaining
data for machine learning experiments.

\section{Logical system}

The logical system of GeoLogic consists of a
\emph{logical core} interacting with \emph{tools}.
The logical core contains the following data.
\begin{itemize}
\item The set of all geometrical objects constructed so far. Every
  object can be accessed as a reference (for logical manipulation), or
  as the numerical object (e.g. coordinates of points, for numerical
  checking).
\item The knowledge database. It consists of a disjoint-set data
structure for equality checking, equation systems for ratios and
angles, and a lookup table for tools.
\end{itemize}
The logical core also possess basic automation techniques for angle and
ratio calculations, and deductions around equality.

A \emph{tool} is a general concept for construction steps, predicates,
or inference rules. It takes a list of geometrical references on an
input (and sometimes additional hyper-parameters), possibly adds some
objects and some knowledge to the logical core and returns a list of
geometrical references on the output, or fails. A tool always fails
if the numerical data do not fit.

Besides that, every tool can be executed in a \emph{check mode}, or in
a \emph{postulate mode}. A tool fails in the check mode (and not in
the postulate mode), if it requires a fact which is not known by the
knowledge database. Otherwise the outcomes of the two modes are the
same

Most tools are memoized. When they are called, their input is
associated with their output in the lookup table of the logical
core. In the next call of the same tool on the same input, the tool
does not fail (even in check mode) and returns the stored output (the
same logical references). This serves three purposes: computation
optimization, functional extensionality and as a database for
predicates.
In particular, a primitive predicate \texttt{lies\_on} is a memoized
tool which in postulate mode only checks whether a given point is
contained by a given line or circle. If it is not, it fails, otherwise
it returns an empty output. In check mode, however, this tool always
fail. It means that the only way how to make this tool executable in
the check mode is to have the input already stored in the lookup
table by calling it in the postulate mode before. Ths differs from
topological (coexact) predicates such as \texttt{not\_on} which in
both modes only checks the numerical conditions -- whether a given
point in not contained by the given line or circle.

By proving a fact (any tool applied to given input) in the logic
system, we mean executing certain tools in the check mode (proof), so
that in the end the given fact can be also run in the check mode.
The graphical interface allows user to run tools in check mode only.

\subsection{Composite tools}
\label{composite-tools}

A composite tool is basically a sequence of other tool steps applied
to the input objects, or on the outputs of prior tools in the
sequence. All composite tools are loaded from an external file, so we
will explain them together with their format. An example code
of the composite tool \texttt{angle} follows.

\begin{verbatim}
angle l0:L l1:L -> alpha:A
  d0 <- direction_of l0
  d1 <- direction_of l1
  alpha <- angle_compute 0 d0 -1 d1 1
\end{verbatim}

The first
line of a composite tool is a header consisting of name, input
objects, forward arrow \texttt{->}, and output objects separated by
space. Every input or output object is given by its label before colon,
and its type after colon. Types are given by letters P (point), L
(line), C (circle), A (angle), D (ratio / dimension). Note that the
format allows name overloading as long as the input types are
different, so there can be an \texttt{angle} tool accepting two lines,
and also another \texttt{angle} tool accepting three points.
Following lines describe the tool steps by output objects, backward
arrow \texttt{<-}, tool name and input objects (possibly with
numerical hyperparameters) separated by
space. Now, we use only labels without types since the parser already knows the
input types and it can infer the output types by the used tool. The
output labels must be unique, unless an anonymous label \texttt{\_} is
used. Among the input parameters, there can be also hyperparameters
in the form of integers, floats, or fractions. It is not relevant how
we mix the hyperparameters with the standard parameters but the order
among hyperparameters, and among parameters matters.

The composite tool we described so far is a \emph{macro} which runs
all its tool steps in the same mode as in what the macro is called. If
any of the steps fails, the entire composite tool fails as well.
Next to macros, there can be \emph{axioms} and \emph{lemmata}.
Axiomatic tool is such a composite tool that contains a single
line \texttt{THEN} among the steps. All the steps after THEN are then
executed in postulate mode, even if the axiomatic tool is called in a
check mode. We call the steps before \texttt{THEN} \emph{assumptions}
and the steps after \texttt{THEN} \emph{implications}.
Axiomatic tools are used for wrapping up primitive
constructions (see \texttt{direction\_of}, and \texttt{line}),
or formulating real axioms (see \texttt{isosceles\_ss}).

\begin{verbatim}
direction_of l:L -> a:A
  THEN
  a <- prim__direction_of l

line A:P B:P -> p:L
  <- not_eq A B
  THEN
  p <- prim__line A B
  <- lies_on A p
  <- lies_on B p

isosceles_ss A:P B:P C:P -> 
  <- not_eq B C
  <- eq_dist A B A C
  THEN
  <- eq_angle A B C B C A
\end{verbatim}

Finally, a \emph{lemma} is similar to the axiomatic tool with the
exception that there is a third sequence of steps
(called \emph{proof}) following a \texttt{PROOF} line. When a lemma is
executed in a check-mode, it works the same as an axiomatic tool, but
it also calls a \emph{proof check}.
Proof check constists of the following steps:
\begin{enumerate}
\item opening a new logical core for the following steps,
\item adding the numerical values of input objects as the initial objects,
\item running the assumptions in postulate mode,
\item running the proof in check mode,
\item running the implications in check mode.
\end{enumerate}
If all the tools succeed, the proof check is considered successful.

\begin{verbatim}
isosceles_aa A:P B:P C:P -> 
  <- not_collinear A B C
  <- eq_angle A B C B C A
  THEN
  <- eq_dist A B A C
  PROOF
  <- sim_aa_r C A B B A C
\end{verbatim}

Adding a macro or a lemma to the tool set creates a conservative
extension of the logic -- anything that is provable with the usage of
lemmata and macros can be proven without them.

\section{Example -- Simson's line}

We provide an example GeoLogic usage on the example of proving Simson's line.
All the Figures in this section are exported from GeoLogic,
demonstrating its visualisation of known facts. The code below
representing the constructions and reasoning steps was created inside
GeoLogic's graphical interface.

We start by drawing a triangle $ABC$, and a point $X$ on its
circumcircle.
\begin{verbatim}
A <- free_point -79.20758056640625 -119.095947265625
B <- free_point -126.97052001953125 23.91351318359375
C <- free_point 108.5352783203125 19.20867919921875
a <- line B C
b <- line C A
c <- line A B
o <- circumcircle A B C
X <- m_point_on 0.6169557687823527 o
\end{verbatim}

\newdimen\imgwidth
\imgwidth=5cm
\centerline{%
  \includegraphics[width = \imgwidth]{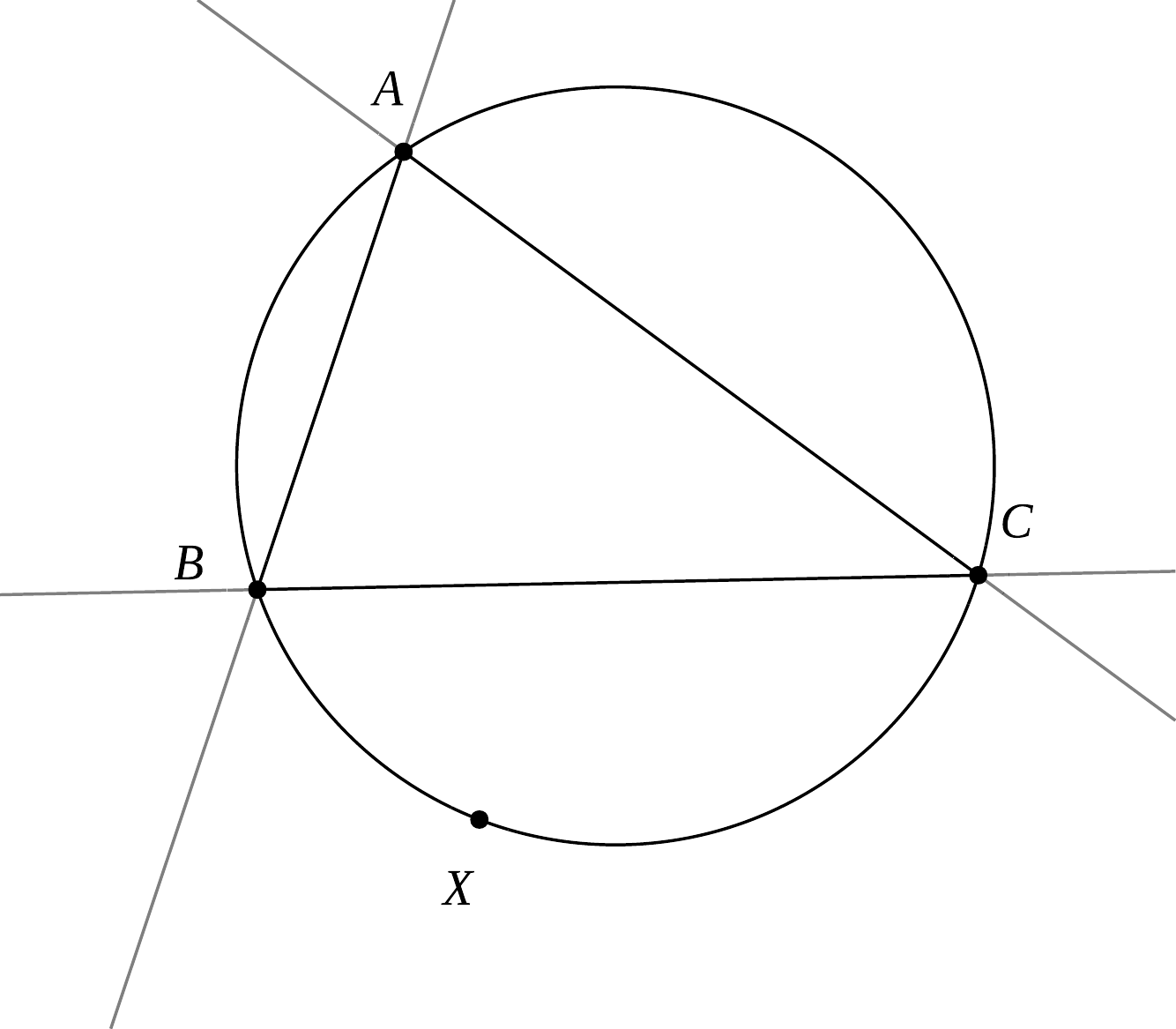}%
}

Simson's line is a line passing through foots $F_a$, $F_b$, $F_c$ of
the point $X$ to the sides of the triangle. However, GeoLogic is not
aware (yet) of the fact that these three points are collinear.

\begin{verbatim}
Fa <- foot X a
Fb <- foot X b
Fc <- foot X c
d <- line Fc Fa
e <- line Fb Fa
\end{verbatim}

\centerline{%
  \includegraphics[width = \imgwidth]{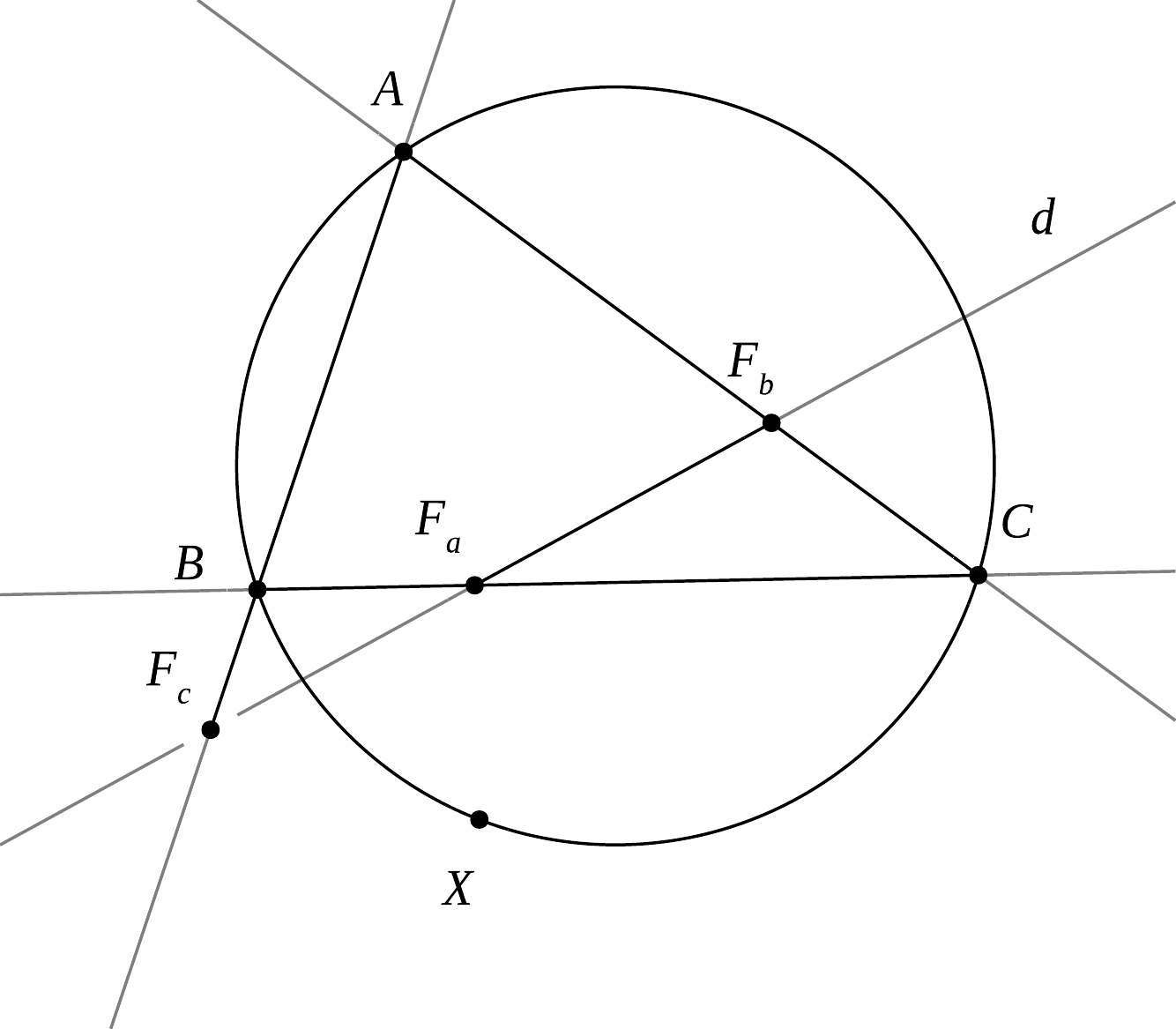}%
  \includegraphics[width = \imgwidth]{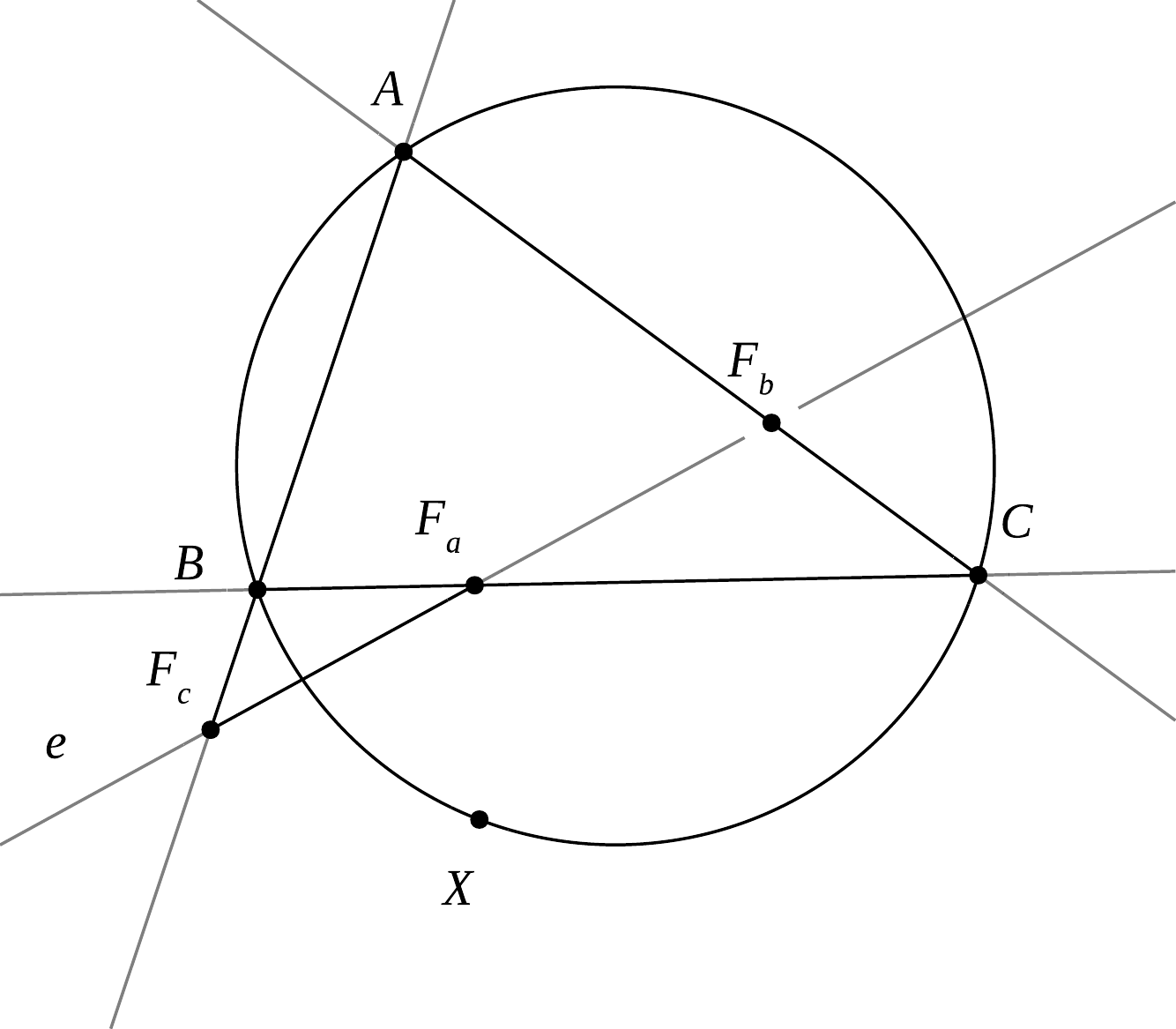}%
}
\bigskip

We can use the fact that the angles $CF_aX$ and $CF_bX$ are equal
(they are both right angles) to conclude that points $C$, $X$,
$F_a$, $F_b$ are concyclic. We consequently use this fact to obtain that
the angles $F_bF_aC$ and $F_bXC$ are equal.

\begin{verbatim}
<- angles_to_concyclic C X Fa Fb
<- concyclic_to_angles Fb C X Fa
\end{verbatim}

We can similarly reason that the points $B$, $X$, $F_a$, $F_c$ are
concyclic and consequently the angles $BF_aF_c$ and $BXF_c$ are
equal.

\begin{verbatim}
<- angles_to_concyclic B X Fc Fa
<- concyclic_to_angles Fc B Fa X
\end{verbatim}

\centerline{%
  \includegraphics[width = \imgwidth]{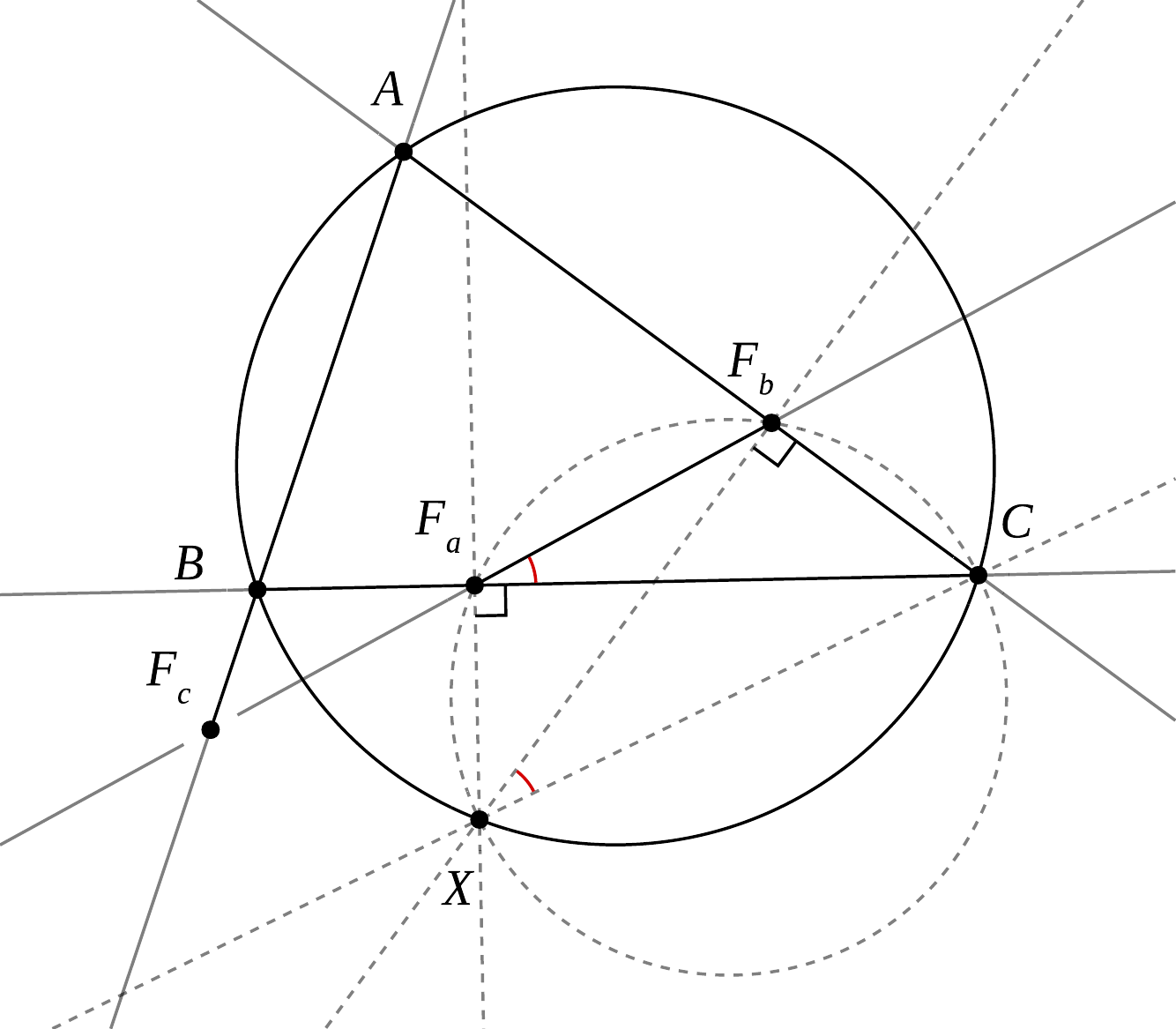}%
  \includegraphics[width = \imgwidth]{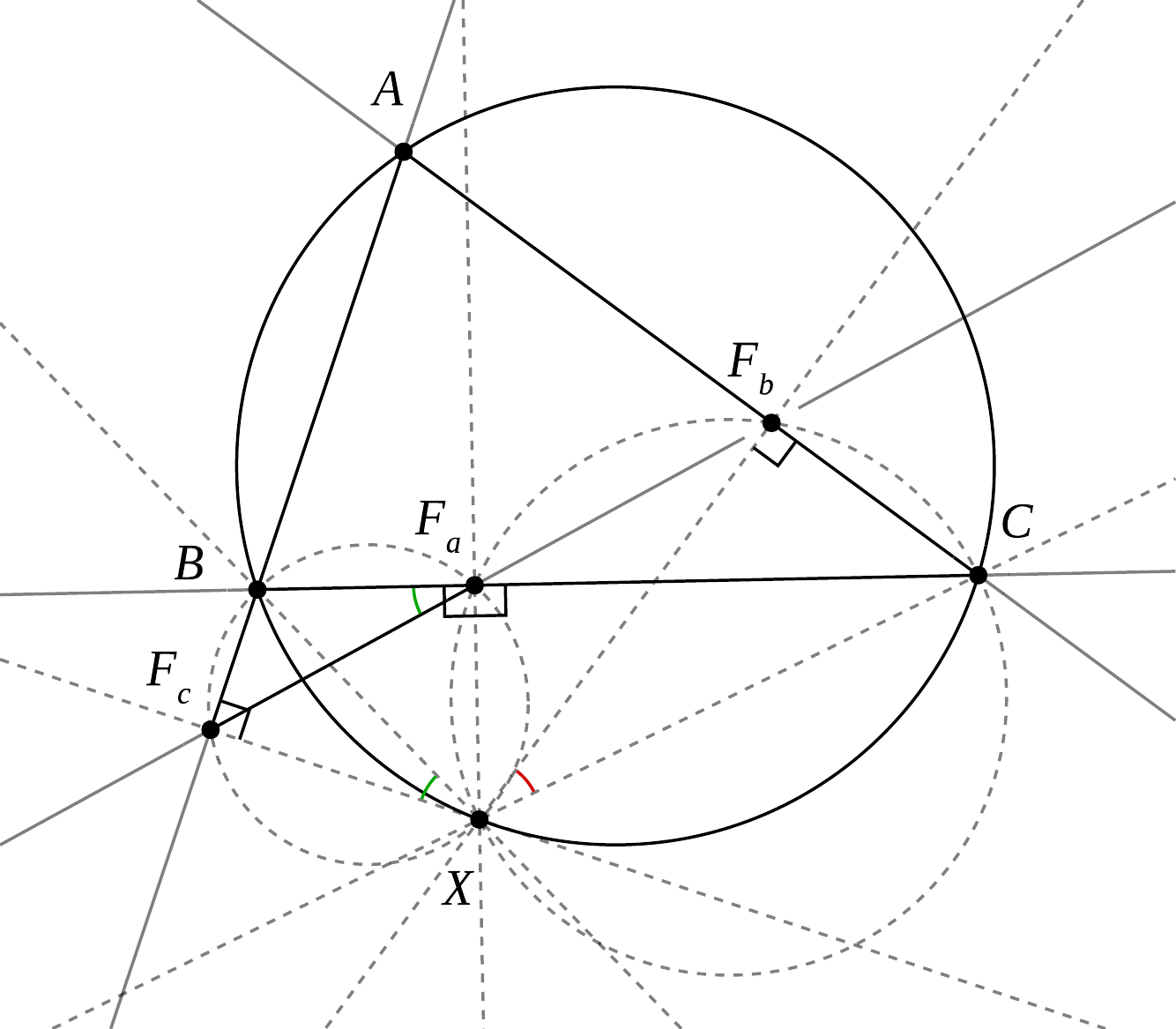}%
}
\bigskip

Finally, we use concyclicity of $X$, $A$, $C$, $B$ to conclude that
the angle $XCA$ is equal to the complementary angle of $ABX$.

\begin{verbatim}
<- concyclic_to_angles X A C B
\end{verbatim}

From this point on, GeoLogic's logical core realizes by itself that
$$
\angle BF_aF_c = \angle BXF_c = 90^\circ-F_cBX = 90^\circ-F_bCX =
CXF_b = CF_aF_b,
$$
and since $BF_aC$ are collinear, $F_cF_aF_b$ are collinear as well.

\bigskip
\centerline{%
  \includegraphics[width = \imgwidth]{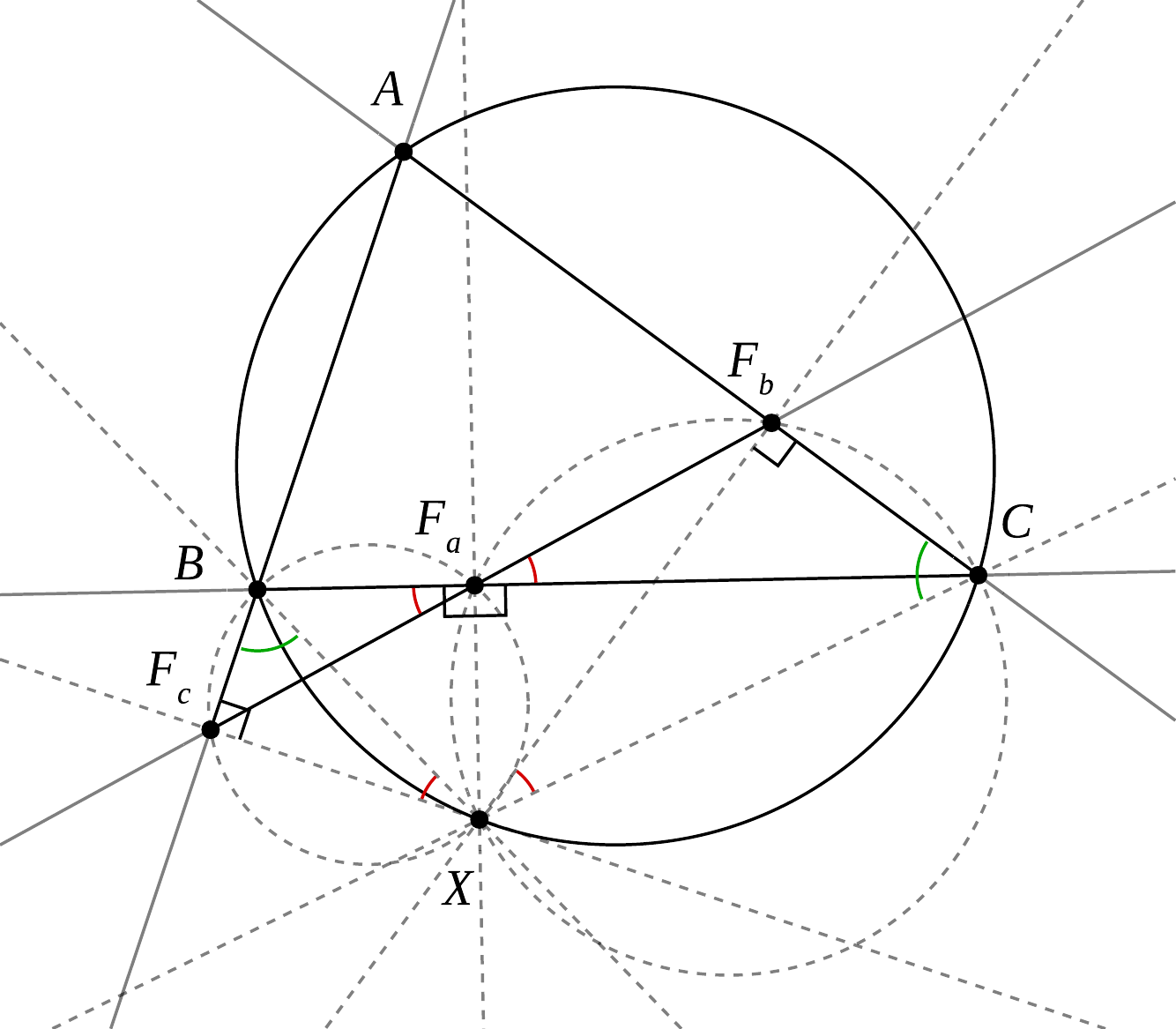}%
}

\section{Related work}

Jeremy Avigad et al.~\cite{Coexact} developed a logical system for
formalizing elementary geometrical proofs from Euclid's elements, also
distinguishing exact and coexact predicates. Their approach is more
formal than ours allowing also proving the coexact statements in the
end but it is less extensible by further tools.
Michael Beeson et al.~\cite{GeoCoQ} connected the interactive theorem
prover CoQ with GeoGebra for visualisation of the theorem (but not for
the proving procedure). Also note that using a rigid logic system such
as in CoQ does not allow numerical checks to be trusted in coexact
statements.

The logical core of GeoLogic is partially inspired by General
Deduction Database~\cite{GDD} and Full Angle~\cite{FullAngle} methods for authomated synthetic proofs in
Euclidean Geometry.
These methods are supported by a graphical application
Geometry Expert~\cite{GEX} which allows user to state a
geometrical problem, run an automated geometrical theorem prover on
it, and visualise the proof.
Julien Narboux presented a similar graphical interface for
construction of geometrical statement traslated to
CoQ~\cite{GeoProof}.
None of these tools, however, supports constructing and checking
proofs in the graphical interface.

\section{Conclusion}

We designed a semi-formal logic for Euclidean geometry which can be to
great extent controlled with a graphical interface and allows us
to prove many standard high school problems. In the future, we would
like to perform experiments with machine learning agents.

\noindent{\bf Acknowledgement.} Supported by the ERC starting grant no.714034 SMART.

\end{document}